\documentclass{aa}
\usepackage{psfig} 
\usepackage{natbib}
\usepackage{graphicx}
\title{A 3-5$\mu$m VLT spectroscopic survey of embedded young low mass stars II$^*$}
\subtitle{Solid OCN$^{-}$}
\author{F.A. van Broekhuizen\inst{1}, K.M. Pontoppidan\inst{2}, H.J. Fraser\inst{1}, and E.F. van Dishoeck\inst{1,2}}
\offprints{F.A. van Broekhuizen,\\e-mail: fvb$@$strw.leidenuniv.nl\\\,$^*$ Based on observations obtained at the European Southern Observatory, Paranal, Chile, within the observing programs 164.I-0605 and 69.C-0441. ISO is an ESA project with instruments funded by ESA Member States (especially the PI countries: France, Germany, The Netherlands and the UK), and with the participation of ISAS and NASA.}
\institute{$^1$ Raymond and Beverly Sackler Laboratory for Astrophysics, Leiden
Observatory, P.O. Box 9513, 2300 RA Leiden, The Netherlands\\$^2$ Leiden Observatory, P.O. Box 9513, 2300 RA Leiden, The Netherlands\\}
\date{31 May 2005, A\&A accepted}
\authorrunning{F.A. van Broekhuizen et al.}
\titlerunning{OCN$^-$....}
\begin{document}
\abstract{The 4.62$\mu$m (2164.5\,cm$^{-1}$) `XCN' band has been
detected in the $M$-band spectra of 34 deeply embedded young stellar
objects (YSO's), observed with high signal-to-noise and high spectral
resolution with the VLT-ISAAC spectrometer, providing the first
opportunity to study the solid OCN$^-$ abundance toward a large number
of low-mass YSO's. It is shown unequivocally that at least two
components, centred at 2165.7\,cm$^{-1}$ (FWHM = 26\,cm$^{-1}$) and
2175.4\,cm$^{-1}$ (FWHM = 15\,cm$^{-1}$), underlie the XCN band. Only
the 2165.7-component can be ascribed to OCN$^-$, embedded in a
strongly hydrogen-bonding, and possibly thermally annealed, ice
environment based on laboratory OCN$^-$ spectra. In order to correct
for the contribution of the 2175.4-component to the XCN band, a
phenomenological decomposition into the 2165.7- and the
2175.4-components is used to fit the full band profile and derive the
OCN$^-$ abundance for each line-of-sight. The same analysis is
performed for 5 high-mass YSO's taken from the ISO-SWS data
archive. Inferred OCN$^-$ abundances are $\leq$\,0.85\,\% toward
low-mass YSO's and $\leq$\,1\,\% toward high-mass YSO's, except for
W33\,A. Abundances are found to vary by at least a factor of 10--20
and large source-to-source abundance variations are observed within
the same star-forming cloud complex on scales down to 400 AU, indicating that the OCN$^-$ formation mechanism is sensitive to local
conditions. The inferred abundances allow quantitatively for
photochemical formation of OCN$^-$, but the large abundance variations
are not easily explained in this scenario unless local radiation
sources or special geometries are invoked.  Surface chemistry should
therefore be considered as an alternative formation mechanism.
 } \maketitle

\section{Introduction}

The 4.62\,$\mu$m (2165 cm$^{-1}$) feature, commonly referred to as the
XCN band, was first detected toward the massive protostar W\,33A by
\citet{Soifer1979}. Its presence was the first observational
indication that complex chemistry could be occurring in interstellar
ice mantles. Subsequently, a similar feature was observed around a
number of other sources, mostly high-mass young stellar objects
(YSO's)
\citep{Tegler1993,Tegler1995,Demyk1998,Pendleton1999,Gibb2000,Keane2001,Whittet2001},
one field star \citep{Tegler1995} and several galactic centre sources
\citep{Chiar2002,Spoon2003}. \\ \indent \citet{Pontoppidan2003}
(henceforth referred to as Paper I) have recently presented $M$-band
spectra of 44 YSO's, 31 of which are low-mass sources
($\leq$\,2\,M$_{\odot}$, $<50$L$_{\odot}$). Most of these sources are
deeply embedded in dark clouds, with few external sources of
ultraviolet radiation, offering a unique opportunity to study the
chemical characteristics of ice mantles in low-mass star-forming
environments. In addition to the strong CO-ice band, observed on most
lines of sight, Paper I reported the detection of a weaker band in 34
of the spectra (27 of which were low-mass objects). This band was
labelled the 2175 cm$^{-1}$ band because its peak-centre position
ranged from 2162-2194 cm$^{-1}$, with an average value of 2175
cm$^{-1}$. These were the first reported detections toward a large
number of low-mass YSO's of a feature similar to the XCN band. \\
\indent The XCN band has been studied extensively in the
laboratory where it is easily reproduced by proton irradiation
\citep{Moore1983}, vacuum ultraviolet photolysis
\citep{Lacy1984,Grim1987,Demyk1998}, or thermal annealing
\citep{Raunier2003,vanBroekhuizen2004} of interstellar ice
analogues. Experimental studies using isotopic substitution proved
unequivocally that the carrier of the laboratory feature is OCN$^{-}$
\citep{Schutte1997,Bernstein2000,Novozamsky2001,Palumbo2000}.  Despite
this assigment, alternative identifications of the interstellar
feature are still debated \citep{Pendleton1999}.
\\ \indent Here we
present a detailed analysis of the 2175 cm$^{-1}$ feature (henceforth
described in this paper as the XCN band) for 34 of the YSO's from
Paper I, plus 5 additional high-mass YSO's taken from the ISO-SWS data
archive, i.e. AFGL\,2136, NGC\,7538 IRS1, RAFGL\,7009S, AFGL\,989 and
W\,33A. Our analysis suggests that two components are required to
reproduce the observational line profiles of the XCN band, only one of
which can be associated with OCN$^-$. The carrier of the second
component is not known but may be due to CO gas-grain interactions, as
described in detail by \citet{Fraser2005}.  
\\ \indent In Sect. 2, the
observational details are summarised, followed by an overview, in
Sect. 3, of laboratory spectroscopy of OCN$^-$ in interstellar ice
analogues and studies of OCN$^-$ formation mechanisms under
interstellar conditions, focusing on UV- and thermally induced OCN$^-$
formation. The fitting methods used for the XCN band analysis are
described in Sect. 4, and the results presented in Sect. 5. In Sect. 6
OCN$^-$ abundances are determined and the implications for the OCN$^-$
formation mechanisms toward the low-mass YSO's studied here are
discussed.
\begin{figure}
\centering
\includegraphics[width=8cm]{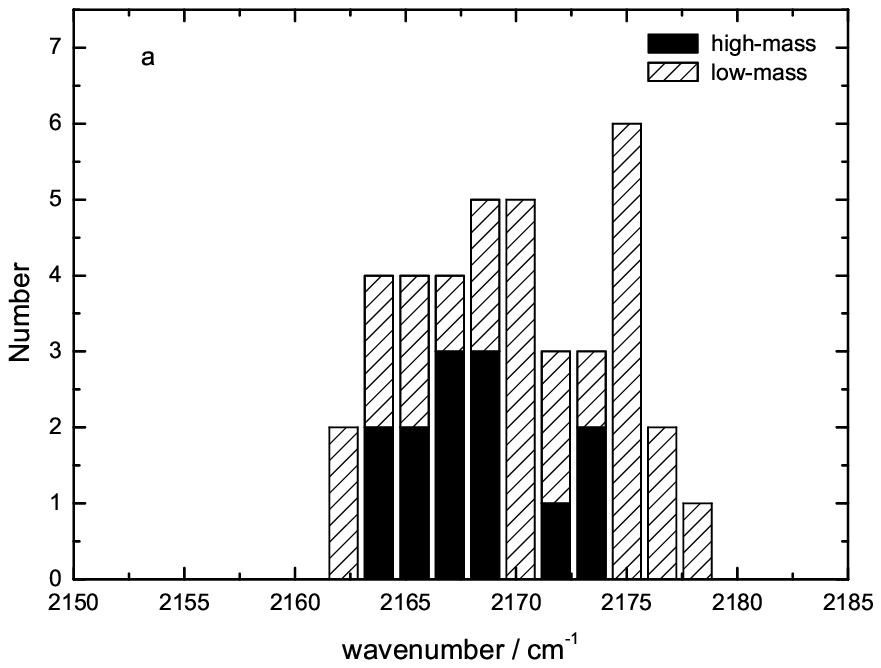}
\includegraphics[width=8cm]{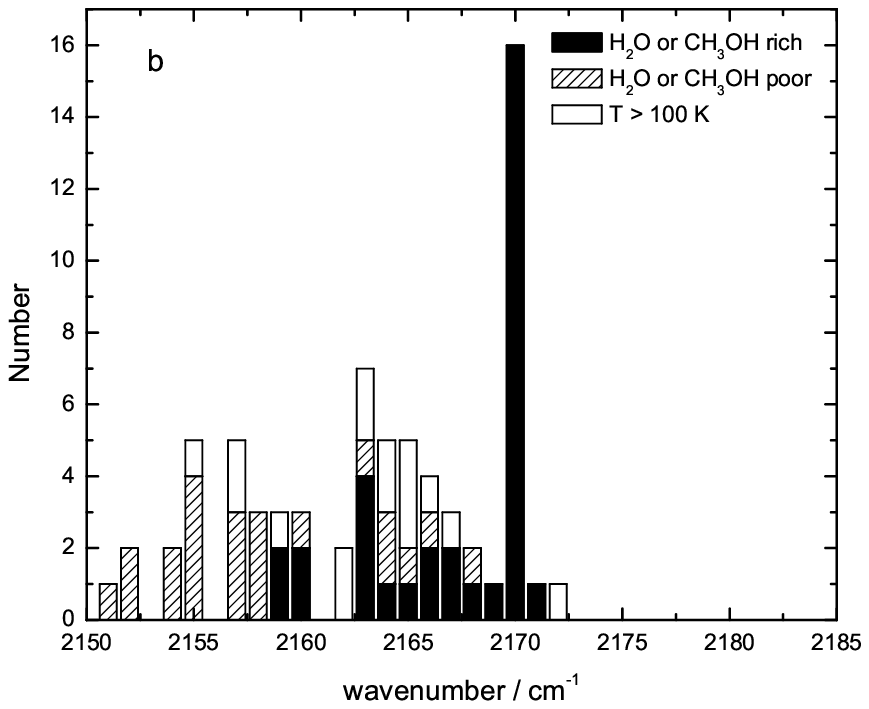}
\caption{{\bf (a)} Observed peak-centre positions of the XCN band toward all 39 low- and high-mass YSO's studied here. Bands originating in high-mass YSO spectra are represented in black, all other YSO's (dominated by low-mass objects) are hatched. Peak centres are binned at 1.6\,cm$^{-1}$. {\bf (b)} Peak-centre positions of laboratory spectra of the $\nu_3$ vibrational band of solid OCN$^-$ in a variety of interstellar ice analogues, also summarised in Tables~\ref{labA} and~\ref{labB}. H$_2$O- and CH$_3$OH-rich ices are shown in black, H$_2$O- and CH$_3$OH-free ices are hatched, and those thermally annealed to $\geq$\,110~K are white. These centre positions are rounded to the nearest integer wavenumber (cm$^{-1}$).}
\label{distr}
\end{figure}
\begin{figure}
\centering
\includegraphics[width=7.6cm]{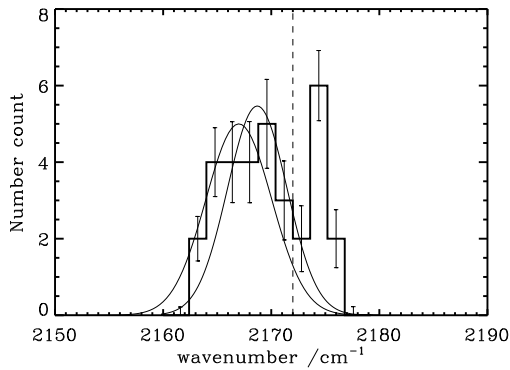}
\caption{The distribution of centre positions of the XCN band
toward the various lines-of-sight in bins of 1.6 cm$^{-1}$ including
the uncertainties derived from the Monte-Carlo simulations.  The
Gaussian at 2168.7\,cm$^{-1}$ is obtained from a single Gaussian fit to
the full data set, that at 2166.9\,cm$^{-1}$ from a similar fit
excluding all XCN band centres $\geq$\,2172\,cm$^{-1}$. This analysis
strengthens the conclusion that two components are present in the observational
data. The dashed line indicates the maximum wavenumber beyond which no
laboratory-based $\nu_3$(OCN$^-$) feature has been found in
interstellar ice analogues.}
\label{monte}
\end{figure}
\section{Observations}
\label{obs}
\subsection{Observational details}
\label{observations}
\indent This work uses the $M$-band spectra of a large sample 
of deeply embedded young stars, which were first
presented in Paper I. The data are part of a 3--5\,$\mu$m
spectroscopic survey of low-mass embedded objects in the nearest
star-forming clouds ($\rho$ Oph, Serpens, Orion, Corona Australis,
Chamaeleon, Vela, and Taurus) using the Infrared Spectrometer And
Array Camera (ISAAC) mounted on UT1-Antu of the Very Large Telescope
(VLT) \citep{vanDishoeck2003}. All sources have spectral energy
distributions representative of Class I objects, with typical
lifetimes of $\sim$\,10$^5$ yr since cloud collapse. The $M$-band
spectra were obtained in the 4.53--4.90\,$\mu$m (2208--2040\,cm$^{-1}$) range using the medium
resolution mode, resulting in a resolving power of
$\lambda$/$\Delta\lambda$ = 5000-10000. The reduction of the spectra
is described in detail in Paper I. In short, the spectra were
corrected for telluric absorption, flux-calibrated relative to bright
early-type standard stars, and wavelength-calibrated relative to
telluric absorption lines in the standard star spectrum with an
accuracy of $\sim$~5 km~s$^{-1}$. The final optical depth spectra were
derived by fitting a blackbody continuum to the 4.52-4.55\,$\mu$m (2212--2198\,cm$^{-1}$) and
4.76-4.80\,$\mu$m (2101--2083\,cm$^{-1}$) regions, where no features are expected, taking care
to exclude known gas-phase lines from the fit. The short wavelength
end of the spectra is more noisy due to the onset of the strong
telluric CO$_2$ features.

In addition, spectra of 5 high-mass sources taken from the ISO-SWS data archive, i.e. AFGL\,2136, NGC\,7538 IRS1, RAFGL\,7009S, AFGL\,989 and W\,33A, were added to the sample.
\subsection{The XCN band}
\label{majorfindings}
\indent In addition to the CO-ice band, a weak secondary feature was
detected in a significant subset of the lines of sight. In Paper I
this feature was reasonably well reproduced using a single Gaussian
whose peak-centre position varied significantly (from 2162 to
2194\,cm$^{-1}$) and Full Width Half Maxima (FWHM) ranged from 9 to
36\,cm$^{-1}$. This distribution is summarised in Fig.~\ref{distr}a,
where the peak-centres of all XCN bands in the observed sample are
plotted, together with the data from the ISO sources.  
\\ \indent Careful inspection of the XCN band shows that in some sources like
W33\,A the feature peaks at 2165.7\,cm$^{-1}$ whereas in other sources
like Elias\,32 and IRS\,63 the feature is shifted much further to the
blue, around 2175\,cm$^{-1}$. This conclusion that two components
underlie the XCN band is supported by a statistical analysis of the
observed centre positions shown in Fig.~\ref{monte}.  The
uncertainties in the observed distribution were quantified using a
Monte Carlo approach.  Each Monte Carlo run varies the XCN band centre
position of each source within a normal distribution with a standard
deviation from peak-centre position as found in Paper I. This produces a
perturbed distribution allowed within the uncertainties on the band
positions. By calculating a large number ($10^5$) of perturbed
distributions, an uncertainty on the number of sources within each
wavenumber bin can be estimated by simply deriving the standard
deviation for the set of perturbed distributions.  A single Gaussian
distribution centred at 2168.7\,$\pm$\,1.9 cm$^{-1}$
($\chi^2$\,=\,5.58) is seen to give a poor fit. When all bands centred
at $\geq$\,2172\,cm$^{-1}$ are excluded from the fit, the centre of
the Gaussian distribution shifts to 2166.9\,$\pm$\,1.5\,cm$^{-1}$
($\chi^2$ = 1.57). The converging $\chi^2$ confirms that two
components, i.e. one centred at around 2166\,cm$^{-1}$ and one at
around 2175\,cm$^{-1}$, provide a much better fit.
\subsection{H$_2$O-ice column densities}
\indent In order to derive the OCN$^-$ abundances, water ice column
densities were estimated using $L$-band spectra (2.85--4.2\,$\mu$m)
from the same survey (Dartois et al., in prep.). Optical depth spectra
of the H$_2$O-ice feature were obtained by fitting a blackbody
spectrum to a $K$-band photometric point from the 2MASS point source
catalogue and the available spectroscopic points in the $L$- and
$M$-band spectra at wavelengths longer than 3.8\,$\mu$m. Care was
taken to exclude any hydrogen recombination lines from the
fit. Estimated uncertainties in the water ice optical depths are
$\sim$\,20\%, mainly due to the variable near-infrared fluxes of
embedded young stars \citep[see e.g.][]{Kaas1999}. The optical depths
were converted to column densities using a factor of
$1.56\times10^{18}\,\rm cm^{-2}$. This factor was derived by
integrating over a laboratory-based H$_2$O spectrum fitted to the
highest quality water bands and scaling with a band strength of
$2\times 10^{-16}\,\rm cm\,molec.^{-1}$ \citep{Gerakines1995}.

\begin{table*}
\centering
\caption{Laboratory experiments on the spectroscopy of $\nu_3$(OCN$^-$) in interstellar ice analogues (T = 10-100\,K)\,$^{a}$}
\label{labA}
\tiny
\begin{tabular}{l|l|l|cc|r}
\hline
\hline
T    & Energetic        & \multicolumn{1}{c|}{Ice composition}                              & $\nu_3$(OCN$^-$) & FWHM           & ref.\\
(K)    & processing       & \multicolumn{1}{c|}{(spectral resolution where known /cm$^{-1}$)}                                             & (cm$^{-1}$)        & (cm$^{-1}$)     & \\
\hline\\
10   & ion (30keV)      & H$_2$O/CH$_4$/NH$_3$ = 5:4:2 (2)                & 2170             &                 &  1\\
\it{10}   & \it{thermal}         & \it{HNCO/NH$_3$ = 1:10 (0.5)}                          & \it{2151}            &                 &  2\\
10   & UV (at 10K)      & H$_2$O/CO/CH$_4$/NH$_3$ = 6:2:1:1 (2)           & 2167             & 25              &  3\\
\it{10}  & \it{UV (at 10K)}     & \it{CO/NH$_3$ = 3:1}                             & \it{2155}            &                 &  4\\
10   & UV (at 10K)      & H$_2$O/CO/CH$_3$OH/NH$_3$ = 100:10:50:10     & 2168             &                 &  5\\
10   & UV (at 10K)      & H$_2$O/CO/CH$_3$OH/C$_n$H$_{2n+2}$/NH$_3$ = 100:10:50:10:10 (0.9)    & 2160             &                 &  6\\
\it{12}  & \it{UV (at 12K)}     & \it{HNCO/NH$_3$ = 1:1 (2)}                           & \it{2155}            &                 &  7\\
\it{12}  & \it{UV (at 12K)}     & \it{HNCO/NH$_3$ = 1:100 (2)}                         & \it{2155}            &                 &  7\\
\it{12 } & \it{UV (at 10K)}     & \it{CO/NH$_3$ = 1:40 (2)}                            & 2152             &                 &  7\\
\it{12}  & \it{UV (at 30K)}     & \it{CO/NH$_3$ = 1:40 (2)}                            & \it{2152}            &                 &  7\\
\it{12}  & \it{ion (60keV)}     & \it{CO$_2$/N$_2$ = 1:1 (1)}                          & \it{2168}            &                 &  8\\
15   & proton (0.8MeV)  & H$_2$O/CO/NH$_3$ = 5:1:1 (4)                     & 2167             & 26              &  9\\
15   & proton (0.8MeV)  & H$_2$O/CO$_2$/NH$_3$ = 1:1:2 (4)                & 2164             & 25              &  9 \\
15   & proton (0.8MeV)  & H$_2$O/CH$_4$/NH$_3$ = 1:1:1 (4)               & 2159             & 23              &  9\\
15   & proton (0.8MeV)  & H$_2$O/CO$_2$/N$_2$ = 5:1:1 (4)                & 2170             & 25              &  9\\
15   & proton (0.8MeV)  & H$_2$O/CH$_4$/N$_2$ = 1:1:1 (4)                 & 2159             & 27              &  9\\
\it{15}  & \it{proton (0.8MeV)} & \it{CO/NH$_3$ = 1:1 (4)}                             & \it{2164}            & \it{27}             &  9\\
\it{15}  & \it{proton (0.8MeV)} & \it{CO$_2$/NH$_3$ = 1:2 (4)}                         & \it{2158 }           & \it{26}             &  9\\
\it{15}  & \it{proton (0.8MeV)} & \it{CO/N$_2$ = 1:1 (4)}                              & \it{2164}            & \it{27}             & 10\\
\it{15}  &\it{ proton (0.8MeV)} & \it{CO/N$_2$ = 1:2 (4)}                              & \it{2158 }           & \it{26}             & 10\\
15   & UV (at 15K)      & H$_2$O/HNCO/NH$_3$ = 120:1:10 (2)               & 2169             & 25              & 11\\
15   & UV (at 15K)      & H$_2$O/HNCO/NH$_3$ = 120:1:10 (2)               & 2171             & 25              & 11\\
\it{15}  & \it{UV (at 15K)}     & \it{HNCO/NH$_3$ = 1:10 (2)}                          & \it{2160 }           &  \it{20}             & 11\\
15   & UV (at 15K)      & H$_2$O/HNCO/NH$_3$ = 140:8:1 (2)                & 2165             & 25              & 11\\
20   & proton (1MeV)    & H$_2$O/CH$_4$/NH$_3$ = 1:2:3                 & 2170             &                 & 12\\
20   & proton (1MeV)    & H$_2$O/CH$_4$/NH$_3$ = 1:2:4                 & 2170             &                 & 12\\
20   & proton (1MeV)    & H$_2$O/CH$_4$/NH$_3$ = 2:4:7                 & 2170             &                 & 12\\
20   & proton (1MeV)    & H$_2$O/CH$_4$/NH$_3$ = 1:7:10                & 2170             &                 & 12\\
20   & proton (1MeV)    & H$_2$O/CH$_4$/NH$_3$ = 1:3:6                 & 2170             &                 & 12\\
20   & proton (1MeV)    & H$_2$O/CH$_4$/NH$_3$ = 1:3:3                 & 2170             &                 & 12\\
20   & proton (1MeV)    & H$_2$O/CH$_4$/NH$_3$ = 1:4:3                 & 2170             &                 & 12\\
20   & proton (1MeV)    & H$_2$O/CH$_4$/NH$_3$ = 15:7:9                & 2170             &                 & 12\\
20   & proton (1MeV)    & H$_2$O/CH$_4$/N$_2$ = 1:1:1                  & 2170             &                 & 12\\
20   & proton (1MeV)    & H$_2$O/CO/N$_2$ = 5:1:1                      & 2170             &                 & 12\\
20   & proton (1MeV)    & H$_2$O/CO/NH$_3$ = 2:2:1                     & 2170             &                 & 12\\
20   & proton (1MeV)    & H$_2$O/CO/NH$_3$ = 2:1:3                     & 2170             &                 & 12\\
20   & proton (1MeV)    & H$_2$O/CO/NH$_3$ = 5:1:10                    & 2170             &                 & 12\\
\it{30}  & \it{UV (at 12K)}     &\it{HNCO/NH$_3$ = 1:100}                         & \it{2155 }           &                 &  7\\
\it{30}  & \it{thermal}         & \it{HNCO/NH$_3$/Ar = 2:2:1000 (0.125)}                  & \it{2157}            &                 &  2\\
30   & UV (at 10K)      & H$_2$O/CO/CH$_3$OH/C$_n$H$_{2n+2}$/NH$_3$ = 100:10:50:10:10  (0.9)   & 2163             &                &  6\\
\it{35}  & \it{proton (0.8 MeV)}& \it{CO/CH$_4$/N$_2$ = 1:1:100  (1)}                   & \it{2166}            & \it{25}             & 13\\
\it{50 } & \it{UV (at 12K)}     &\it{HNCO/NH$_3$ = 1:100 (2)}                         & \it{2154}            &                 &  7\\
\it{50 } & \it{UV (at 37K)}     & \it{CO/NH$_3$ = 1:1 (2)}                             & \it{2158}            &                 &  7\\
55   & UV (at 10K)      & H$_2$O/CO/CH$_3$OH/C$_n$H$_{2n+2}$/NH$_3$ = 100:10:50:10:10  (0.9)   & 2163             &                 &  6\\
\it{60}  & \it{UV (at 12K)}     & \it{HNCO/NH$_3$ = 1:100 (2)}                         & \it{2154}            &                 &  7\\
\it{80}  & \it{UV (at 12K)}     & \it{CO/NH$_3$ = 1:1 (4)}                             & \it{2157}            &                 & 14 \\
80   & UV (at 12K)      & H$_2$O/CO/NH$_3$ = 1:1:1 (4)                    & 2160             &                 & 14\\
80   & UV (at 12K)      & H$_2$O/CO/NH$_3$ = 2:1:1 (4)                     & 2163             &                 & 14\\
80   & UV (at 12K)      & H$_2$O/CO/NH$_3$ = 3:1:1 (4)                    & 2166             &                 & 14\\
\it{100} & \it{UV (at 37K)}     & \it{CO/NH$_3$ = 1:1}                             & \it{2163}            &                 & 15\\
\it{100 }& \it{UV (at 37K)}     & \it{CO/NH$_3$ = 1:1 (2)}                             & \it{2157}            &                 &  7\\
\it{100} & \it{ion (60keV)}     & \it{CO$_2$/N$_2$ = 1:1 (1)}                          & \it{2165}            &                 &  8\\
100  & UV (at 10K)      & H$_2$O/CO/CH$_3$OH/NH$_3$ = 100:10:50:10     & 2166             &                 &  5\\
100  & UV (at 10K)      & H$_2$O/CO/CH$_3$OH/C$_n$H$_{2n+2}$/NH$_3$ = 100:10:50:10:10  (0.9)   & 2163             &                 &  6\\
\hline
\end{tabular}
\begin{itemize}
\item[]\footnotesize{$^{a}$Experiments in italics show ices that lack H$_2$O. All $\nu_3$ peak-positions are rounded up or down to the nearest integer wavenumber. $^1$ \citet{Strazzulla1998}, $^2$ \citet{Raunier2003}, $^3$ \citet{Dhendecourt1986}, $^4$ \citet{Lacy1984}, $^5$ \citet{Tegler1993}, $^6$ \citet{Allamandola1988}, $^7$ \citet{Novozamsky2001}, $^8$ \citet{Palumbo2000}, $^9$ \citet{Hudson2000}, $^{10}$ \citet{Hudson2001}, $^{11}$ \citet{vanBroekhuizen2004}, $^{12}$ \citet{Moore1983}, $^{13}$ \citet{Moore2003}, $^{14}$ \citet{Grim1987}, $^{15}$ \citet{Bernstein2000}}
\end{itemize}
\end{table*}
\begin{table*}
\centering
\caption{Laboratory experiments on the spectroscopy of OCN$^-$ in interstellar ice analogues (T\,$>$~100K)\,$^{a}$}
\label{labB}
\tiny
\begin{tabular}{l|l|l|cc|r}
\hline
\hline
T    & Energetic        & \multicolumn{1}{c|}{Ice composition}                              & $\nu_3$(OCN$^-$) & FWHM            & ref.\\
(K)    & processing       &  \multicolumn{1}{c|}{(spectral resolution where known /cm$^{-1}$)}                                            & (cm$^{-1}$)        & (cm$^{-1}$)     & \\
\hline\\
110  & thermal          & H$_2$O/HNCO/Ar = 12:1:1000 (0.12)                  & 2170             &                 & 1\\
\it{120}  & \it{UV (at 10K)}      &\it{CO/NH$_3$ = 1:1}                              & \it{2155}             &                 & 2\\
120  & UV (at 15K)      & H$_2$O/HNCO/NH$_3$ = 120:1:10 (2)               & 2166             & 22              & 3\\
130  & thermal          & H$_2$O/HNCO = 10:1 (0.5)                          & 2172             &                 & 1\\
\it{140}  & \it{UV (at 37K)}      & \it{CO/NH$_3$ = 1:1  (2) }                            & \it{2159}             &                 &  4\\
\it{150}  & \it{ion (60keV)}      & \it{CO$_2$/N$_2$ = 1:1 (1)}                           & \it{2163}             &                 &  5\\
\it{150}  & \it{UV (at 10K)}      & \it{CO/NH$_3$ = 3:1}                              & \it{2165}             &                 &  6\\
150  & UV (at 10K)      & H$_2$O/CO/CH$_3$OH/NH$_3$ = 100:10:50:10     & 2162             &                 &  7\\
\it{160}  & \it{thermal}          & \it{HNCO/NH$_3$ = 1:10 (0.5)}                          & \it{2165}             &                 &  1\\
\it{180}  & \it{UV (at 12K)}      & \it{CO/NH$_3$ = 1:1 (4)}                             & 2157             &                 &  8\\
\it{180}  & \it{UV (at 12K)}      & \it{CO/NH$_3$ = 1:10 (4)}                            & \it{2157}             &                 &  8\\
\it{200}  & \it{ion (60keV)}      &\it{CO$_2$/N$_2$ = 1:1 (1)}                          &\it{2162}             &                 &  5\\
200  & UV (at 10K)      & H$_2$O/CO/CH$_4$/NH$_3$ = 6:2:1:1 (2)           & 2167             & 25              &  9\\
200  & UV (at 10K)      & H$_2$O/CO/CH$_3$OH/NH$_3$ = 100:10:50:10     & 2164             &                 &  7\\
200  & UV (at 10K)      & H$_2$O/CO/CH$_3$OH/NH$_3$ = 100:10:50:10 (0.9)    & 2165             &                 &  10\\
200  & UV (at 10K)      & H$_2$O/CO/CH$_3$OH/C$_n$H$_{2n+2}$/NH$_3$ = 100:10:50:10:10 (0.9)    & 2164             &                 &  10\\
200  & UV (at 10K)      & H$_2$O/CO/CH$_3$OH/C$_n$H$_{2n+2}$/NH$_3$ = 100:10:50:10:10 (0.9)    & 2163             &                 &  10\\
\hline
\end{tabular}
\begin{itemize}
\item[]\footnotesize{$^{a}$Experiments in italics show ices that lack H$_2$O. All $\nu_3$ peak-positions are rounded up or down to the nearest integer wavenumber. $^1$ \citet{Raunier2003}, $^2$ \citet{Schutte1997}, $^3$ \citet{vanBroekhuizen2004}, $^4$ \citet{Novozamsky2001}, $^5$ \citet{Palumbo2000}, $^6$ \citet{Lacy1984}, $^7$ \citet{Tegler1993}, $^8$ \citet{Grim1987} $^9$ \citet{Dhendecourt1986}, $^{10}$  \citet{Allamandola1988}}
\end{itemize}
\end{table*}
\begin{figure}
\centering
\includegraphics[width=7.1cm]{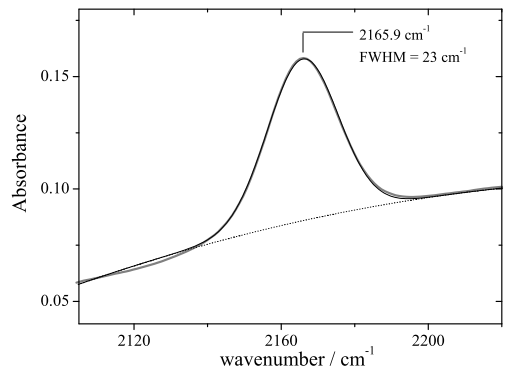}
\caption{Example of a laboratory spectrum of $\nu_3$(OCN$^-$), obtained after thermal annealing to 120\,K of an ice composed of H$_2$O/HNCO/NH$_3$ = 120/1/10 (grey curve). After accounting for the baseline (dotted line), an excellent fit to the laboratory data is obtained with a single Gaussian fit, centred at 2165.9\,cm$^{-1}$, FWHM\,=\,23\,cm$^{-1}$ (black curve).}
\label{ocngauss}
\end{figure}
\begin{figure}[ht!]
\centering
\includegraphics[width=8.4cm]{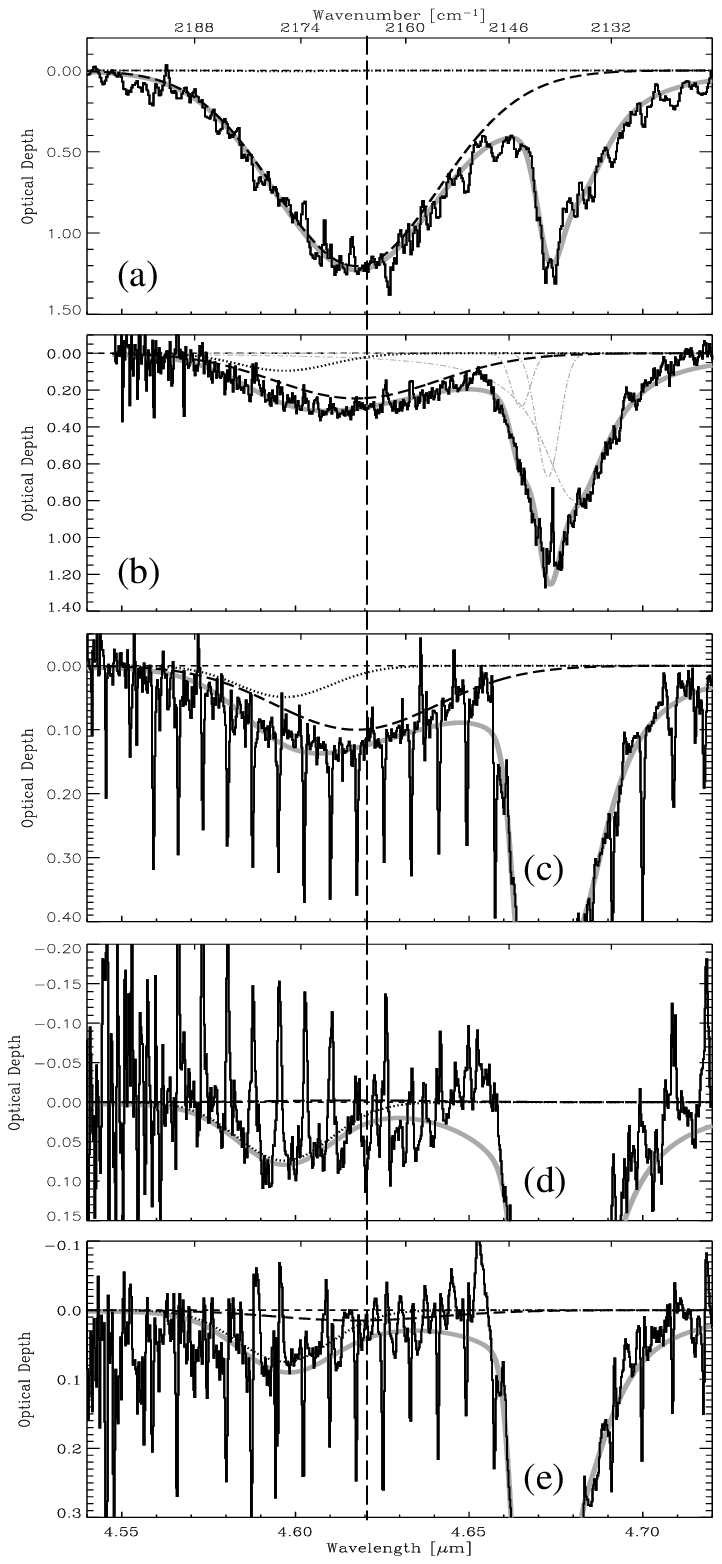}
\caption{The XCN band as observed in the line-of-sight toward five YSO's, ordered from top to bottom from the most red-centred to the most blue-centred band, i.e. {\bf (a)} W33\,A, {\bf (b)} HH\,46, {\bf (c)} Reipurth\,50, {\bf (d)} Elias\,32 and {\bf (e)} IRS\,63. All spectra (black) are continuum subtracted (small-dashed line) and the total fit is shown (grey curve). The two components underlying the XCN band, i.e. the 2165.7-component (dashed curve) and the 2175.4-component (dotted curve) are shown. The three-component fit to the solid CO band at 4.67\,$\mu$m (2139\,cm$^{-1}$) is shown only in {\bf (b)} by the three grey dash-dotted curves. The vertical dashed line at 4.62\,$\mu$m guides the eye.}
\label{fitobs}
\end{figure}
\section{Laboratory experiments on OCN$^-$ pertinent to the XCN band analysis}
\label{labspec}
\subsection{Spectroscopy}
\label{ocnspec}
\indent The formation and infrared spectroscopy of solid OCN$^-$ has been studied extensively in numerous laboratories. Its spectrum is dominated by the CN stretching-vibration (the $\nu_3$ band), which can be well fitted by a single Gaussian profile (see for example Fig.~\ref{ocngauss}). In Tables~\ref{labA} and~\ref{labB} an overview is given of all (to the best of our knowledge) published peak centres and FWHM of the $\nu_3$ band of OCN$^-$, as measured in the laboratory under conditions relevant to the interstellar case, e.g. matrix-isolation studies have been excluded. In addition, information is tabulated on the initial ice composition, ice temperature and processing mechanism applied to the ice. With the exception of H$_2$O, CH$_3$OH and NH$_3$, most of the other ice components have fully desorbed by 120 K under high-vacuum laboratory conditions but OCN$^-$ remains in the solid state, stabilised by a counter ion \citep[possibly NH$_4^+$,][]{Demyk1998,Taban2003}. Hence, Table~\ref{labA} contains those experiments in which the ice matrix is predominantly intact, whereas Table~\ref{labB} summarises those in which most ice constituents have desorbed or are desorbing. \\
\indent Fig.~\ref{distr}b shows the distribution of peak centres of the laboratory $\nu_3$ of the OCN$^-$ feature highlighting the initial ice environment, i.e. H$_2$O- or CH$_3$OH-rich (black), H$_2$O- and CH$_3$OH-lacking (hatched) and thermally annealed, i.e. those listed in Table~\ref{labB} ($T$\,$\geq$\,110\,K, white). It is important to note that this histogram is significantly weighted by the parameter space investigated in the laboratory. Peak centres vary strongly with ice composition, ranging from 2151 to 2172\,cm$^{-1}$. Within this range, $\nu_3$(OCN$^-$) peaks at between 2151--2160 cm$^{-1}$ in the absence of H$_2$O and CH$_3$OH \citep{Novozamsky2001,Raunier2003,Grim1987,Hudson2000}, or at between 2160--2172 cm$^{-1}$ when associated with a strongly hydrogen-bonding environment, i.e. H$_2$O- or CH$_3$OH-rich ices \citep{Grim1987}. No trend is apparent in the data relating the actual formation mechanism of OCN$^-$ to its band position. \\
\indent From Fig.~\ref{distr}b is also clear that in warm ices ($T\geq$\,110\,K), $\nu_3$(OCN$^-$) peaks at similar positions to the colder ices. However within a single ice matrix, thermal annealing beyond 100\,K tends to shift the $\nu_3$(OCN$^-$) peak to between 2155 and 2167\,cm$^{-1}$ when OCN$^-$ is stabilised by NH$_4^+$ \citep{Novozamsky2001,Palumbo2000}. In the exclusive presence of H$_2$O (i.e. for H$_2$O/HNCO and H$_2$O/HNCO/Ar), however, $\nu_3$(OCN$^-$) must be stabilised by water solvation (most plausibly H$_3$O$^+$) and seems thermally unaffected, peaking at 2170 to 2172\,cm$^{-1}$. \\
\indent The spectral range of peak positions observed for laboratory $\nu_3$(OCN$^-$) implies that only one component of the observed XCN-bands can be associated with OCN$^-$. No laboratory $\nu_3$(OCN$^-$) spectrum in Fig.~\ref{distr}b peaks beyond 2172\,cm$^{-1}$, whereas a significant fraction of the astronomical spectra in Fig.~\ref{distr}a peaks around 2175\,cm$^{-1}$. This is emphasised by the dashed line in Fig.~\ref{monte}, indicating that only those XCN-bands represented by the Gaussian fit to the left of the dashed line can be matched to a laboratory spectrum of OCN$^-$. The peak centres of these XCN-bands can be explained by OCN$^-$ residing in a strong hydrogen-bonding ice environment that may have been thermally annealed up to 200\,K. OCN$^-$ embedded in an ice that lacks a strong hydrogen-bonding network generally peaks red-wards of the XCN-bands observed, even after thermal annealing, causing this kind of ice to be an unlikely interstellar OCN$^-$ environment.  
\subsection{Band strength}
\label{bandstrength}
\indent The reported value of the $\nu_3$(OCN$^-$) band strength, $A_{\rm OCN^-}$, varies considerably in the literature. \citet{Lacy1984} assumed $A_{\rm OCN^-}$\,$\leq$\,1.0$\times$10$^{-17}$ cm\,molec.$^{-1}$, based on the assumption that the cross section of a typical CN-stretching vibration is approximately 5 times less than that of a CO-stretch. \citet{Dhendecourt1986} derived $A_{\rm OCN^-}$ from photolysis experiments to be 1.8--3.6$\times$10$^{-17}$ cm\,molec.$^{-1}$, assuming a constant carbon budget per experiment, followed by \citet{Demyk1998}, who assumed a constant oxygen budget to obtain $A_{\rm OCN^-}$\,$\geq$\,4.3$\times$10$^{-17}$ cm\,molec.$^{-1}$. Both studies probably underestimated the value of $A_{\rm OCN^-}$ due to the formation of infrared-weak or inactive photoproducts, and the neglect of photodesorption of volatile species like CO and CO$_2$ (van Broekhuizen et al. in prep). \\
\indent More recently, \citet{vanBroekhuizen2004} determined $A_{\rm OCN^-}$ from the thermal deprotonation of HNCO by NH$_3$, using the NH$_3$ to NH$_4^+$ conversion as reference, giving $A_{\rm OCN^-}$\,=\,1.3$\times$10$^{-16}$\,cm\,molec.$^{-1}$. In the present analysis of the XCN band, this value of $A_{\rm OCN^-}$ has been adopted because its determination is not influenced by any photodesorption or the formation of photo-products (both known and unknown), which may have affected the previously derived values. The adopted band strength does not affect any subsequent conclusions in this paper on the photo-production of OCN$^-$, since these reaction efficiencies were derived using the same value of the band strength. However, the adopted $A_{\rm OCN^-}$ value does impact on the conclusions drawn if OCN$^-$ is thermally formed, because significantly more HNCO will be required in the solid state if a smaller value of $A_{\rm OCN^-}$ were used.
\subsection{Formation mechanisms}
\label{ocnform}
\indent The presence of OCN$^-$ in interstellar ices has become widely cited as a probe of energetic processing of the protostellar environment, in particular by ultraviolet (UV) photons. However, this assumption has been challenged on the basis of a quantitative set of laboratory experiments studying the efficiency of OCN$^-$ photoproduction, and the possibilities of forming OCN$^-$ thermally \citep{Raunier2003,vanBroekhuizen2004}. To assist the reader in following the subsequent discussions and conclusions of this paper, these two OCN$^-$ formation mechanisms, and the regimes in which they apply, are summarised below.\\
\indent In the laboratory, photochemical processes lead to the formation of
OCN$^-$ at abundances of (at most) 1.9\,\% with respect to H$_2$O-ice,
starting from an interstellar ice analogue containing H$_2$O, CO and
NH$_3$ \citep{vanBroekhuizen2004}. The initial amount of NH$_3$ in these experiments ranges from 8 to 40\% with respect to H$_2$O ice. The abundance of interstellar
NH$_3$ is uncertain, but is observed to be $<$5\% toward W33\,A \citep{Taban2003}. Including the constraint that the abundance of NH$_3$ ice after photolysis cannot be more than 5\% lowers the maximum OCN$^-$ photoproduction yield to 1.2\,\% with respect to H$_2$O ice \citep[see, for
example, Figure 6 of][]{vanBroekhuizen2004}.\\
\indent From these experiments, the maximum
interstellar OCN$^-$ photoproduction yield was constrained to an
abundance of 1.2\,\% with respect to H$_2$O-ice by assuming a maximum
abundance for solid NH$_3$ of 5\,\% with respect to H$_2$O-ice, as
observed toward W\,33A \citep{Taban2003}. If the cosmic ray induced
UV-field
\citep[1.4$\times$10$^3$\,photons\,cm$^{-2}$\,s$^{-1}$,][]{Prasad1983}
is assumed to be the only source of UV-photons, this maximum
1.2\,\% OCN$^-$ abundance would be reached after a UV-fluence
equivalent to a molecular cloud lifetime of 4$\times$10$^8$
years. This is long compared to the assumed age of the YSO's studied
in this paper ($\sim$10$^5$ yr), even if a long pre-stellar phase of
$\sim$\,10$^7$\, yr is included.  \\
\indent Alternatively, laboratory studies show that OCN$^-$ can be formed via
thermal heating of ices containing HNCO in the presence of NH$_3$ or
H$_2$O with an efficiency of 15-100\,\%
\citep{Demyk1998,Raunier2003,vanBroekhuizen2004}. Laboratory
experiments and theoretical calculations show that this solvation
process is even efficient at 15\,K provided that HNCO is sufficiently
diluted in NH$_3$ or H$_2$O-ice \citep{Raunier2003,Park2004}, for the
proton transfer to occur.
\begin{table}
\small
\caption{Components adopted in the fit of the observed XCN bands}
\label{method}
\begin{tabular}{l|l|l}
\hline
\hline
                          &   2165.7-component    & 2175.4-component\\
\hline
Line profile                   &   Gaussian            & Gaussian  \\ 
Centre (cm$^{-1}$)        &   2165.7              & 2175.4  \\     
FWHM (cm$^{-1}$)          &   26                  & 15     \\
\hline
\end{tabular}
\end{table}
\section{Decomposition of the XCN profile}
\label{profiledecomposition}
Our analysis of the XCN band, containing two components (see
Sect.~\ref{majorfindings}), refines the results from Paper I where a
single component fit was used. The component centred at
$\sim$\,2175\,cm$^{-1}$ has been discussed in detail by
\citet{Fraser2005} in terms of gas-surface interactions of CO with
interstellar grains. Thus, in the remainder of this paper the
discussion is focused on the carrier of the component of the XCN band
centred at $\sim$2166\,cm$^{-1}$, and simply uses a two
component fit to correct for any contribution from the component
centred at $\sim$\,2175\,cm$^{-1}$. \\
\indent The distribution of the $\sim$2166\,cm$^{-1}$ component of the
XCN band shows a broad spread of band centres, most probably related
to source-to-source differences in the local ice-environment of the
carrier of this component in the various lines-of-sight. A good
`prototype' for this component is the XCN band of W33\,A, as it is
also one of the most red-centred and strongest of the objects
studied. It matches a Gaussian profile, peaking at
2165.7\,$\pm$\,0.1\,cm$^{-1}$ (FWHM = 26\,$\pm$\,1\,cm$^{-1}$), that
is well reproduced by laboratory spectra of OCN$^-$ (see
Fig.~\ref{ocngauss}). An initial analysis showed that this Gaussian
profile also reproduces the red wing and component of the XCN band of
all other YSO's in this study. Consequently, the XCN band was analysed
using a similar procedure to that employed for the 6.8\,$\mu$m band by
\citet{Keane2001}, whereby the 6.8\,$\mu$m band was also decomposed by
using the most extreme source in the sample, Mon\,R2:IRS3, to fit all
lines-of-sight together with a `rest-spectrum' obtained after the
initial fitting of the other sources. Here, the Gaussian fit to the
W33\,A XCN band is adopted (henceforth referred to as the
2165.7-component), and not the actual YSO spectrum, to avoid any
contribution from the CO-ice band to the fit. \\
\indent The component underlying the distribution of XCN bands centred
at $\sim$2175\,cm$^{-1}$ in Fig.~\ref{monte} was determined
empirically, in effect by fitting the residual of the XCN band after
spectral subtraction of the contribution of the 2165.7-component. This
resulted in a Gaussian, centred at 2175.4\,$\pm$\,3\,cm$^{-1}$ (FWHM =
15\,$\pm$\,6\,cm$^{-1}$). This feature will be referred to as the
2175.4-component. Adopting a different Gaussian for the
2165.7-component, i.e. one centred at 2166.9 or 2168.7\,cm$^{-1}$,
does not significantly affect these fit results, which indicates that
the range of environments observed for the 2165.7-component do not
introduce an error on the fit.\\
\indent The fitting routine is based
on the same IDL algorithms developed in Paper I. All spectra were
fitted using the same 2165.7- and 2175.4-components, keeping the
centre positions and FWHM constant. The optical depths of the two
components are the only free parameters (see Table~\ref{method} for a
summary). In addition, a full fit of the CO-ice feature was run in
parallel to correct for possible overlap between the wing of the
CO-ice feature and the XCN band (see Paper I and Fig.~4b for 
details on the CO-ice fit). 
\section{Results}
\label{results}
\begin{figure}
\centering
\includegraphics[width=8.5cm]{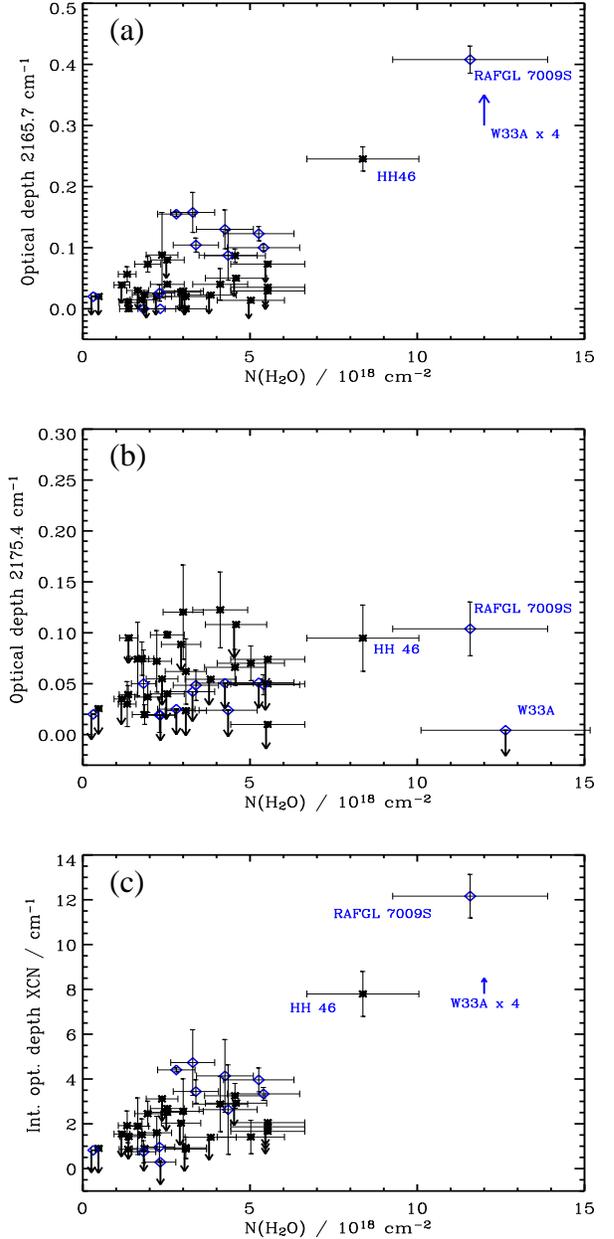}
\caption{{\bf (a)} The optical depth of the 2165.7-component, {\bf (b)} the optical depth of the 2175.4-component, and {\bf (c)} the total integrated area of the XCN band, as determined by Eq.1, are shown versus the H$_2$O ice column density, $N$(H$_2$O). High-mass objects are indicated by diamonds, all others, dominated by low-mass YSO's, by stars. 3\,$\sigma$ uncertainties are given, or upper limits are indicated by arrows. }
\label{h2o4}
\end{figure}
\subsection{Results of the fitting procedure}
Fig.~\ref{fitobs} shows the results of fitting the XCN band in five of
the YSO's in our sample. These are presented from top to bottom in
order of decreasing wavelength of the XCN band centre position. The
XCN band of W33\,A (Fig.~\ref{fitobs}a) shows the excellent fit of the
2165.7-component with $\leq$\,1\,\% contribution by the
2175.4-component. In Fig.~\ref{fitobs}b, already some
contribution to this band from the 2175.4-component is apparent. 
From Fig.~\ref{fitobs}b to e, the
2175.4-component contribution increases, blue-shifting the XCN band,
until it dominates the XCN band toward Elias\,32 and
IRS\,63. Table~\ref{fitresults} summarises the derived optical
depths. The $\chi^2$ of the fit (not shown) generally varies between
0.3 and 10. The
additional three-component fit of the solid CO feature (Paper\,I)
is shown by the
three grey dash-dotted curves in Fig.~\ref{fitobs}b only, but was
applied to all fits. 
\subsection{Correlations with H$_2$O}
\label{corh2o}
In Fig.~\ref{h2o4}a, the optical depth of the 2165.7-component is
plotted against the H$_2$O column density, $N$(H$_2$O). For all YSO's
studied here, the derived optical depth of the 2165.7-component is
$\leq$\,0.16, except for HH\,46, RAFGL\,7009S and W33\,A. No
relationship with $N$(H$_2$O) is apparent. In fact, it seems that the
optical depth value of the 2165.7-component toward high-mass YSO's is
in most cases larger than toward lower mass YSO's, where the optical
depth is typically $\leq$\,0.09.\\ 
\indent Fig.~\ref{h2o4}b presents
the optical depth of the 2175.4-component plotted against
$N$(H$_2$O). The optical depth of the 2175.4-component is
$\leq$0.14 for all sources in this study and is not correlated to the
H$_2$O column density. An interpretation of this feature is given by
\citet{Fraser2005}.\\ \indent Comparing Figs.~\ref{h2o4}a and
\ref{h2o4}b shows that the 2165.7-component dominates the XCN band
toward most high-mass YSO's, whereas the 2175.4-component contributes
most to the XCN band in all other sources. Checks of a possible
relationship between the optical depth of the two components (not
shown) proved negative, indicating that the components do not trace
the same feature.\\ \indent In Fig.~\ref{h2o4}c the total integrated
area of the XCN band is plotted against $N$(H$_2$O). The total
integrated area of the XCN band is derived from the sum of the
integrated areas of the individual components
\begin{equation}
{\int{\tau_{\rm (XCN)}\,d\nu}}={\int{\tau_{2165.7}\,d\nu}}+{\int{\tau_{2175.4}\,d\nu}}
\end{equation}
The XCN band is evenly distributed toward high- and low-mass YSO's and its total area tentatively increases with increasing $N$(H$_2$O). 
\subsection{Correlations with CO}
\label{corco}
\begin{figure}
\centering
\includegraphics[width=8.5cm]{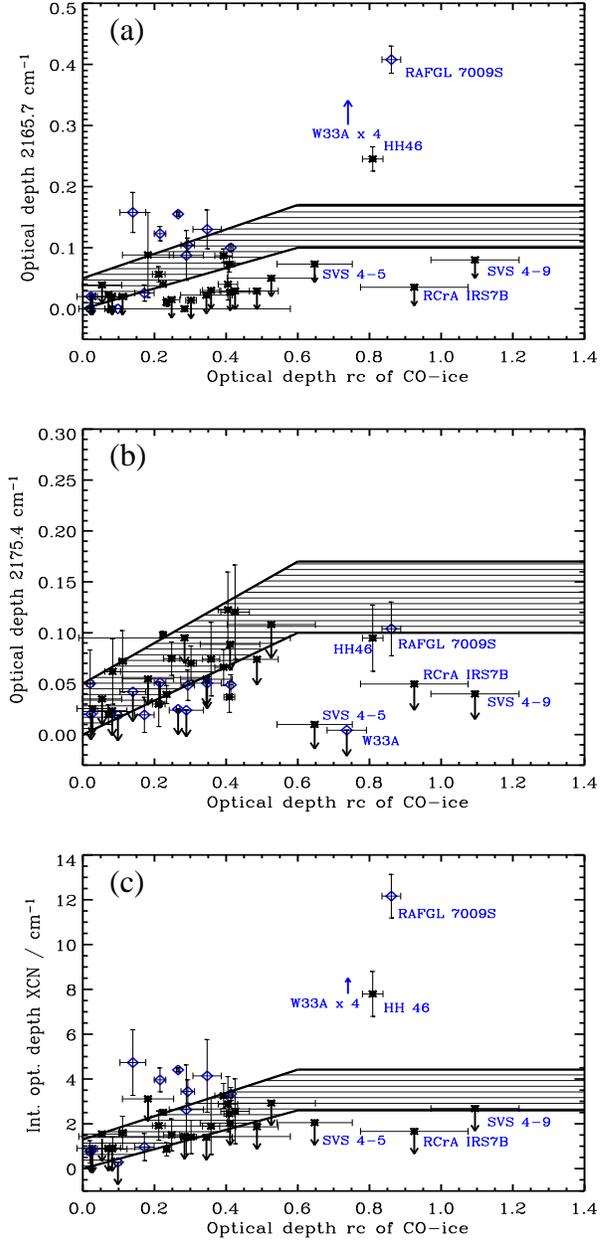}
\caption{{\bf (a)} The optical depth of the 2165.7-component, {\bf (b)} the optical depth of the 2175.4-component, and {\bf (c)} the total integrated area of the XCN band, as determined by Eq.1, are presented versus the optical depth of the red component (rc) of CO-ice. High-mass objects are indicated by diamonds, all others, dominated by low-mass YSO's, by stars. 3\,$\sigma$ uncertainties are given, or upper limits are indicated by arrows. The shaded area, shown in each plot, marks the correlation between the optical depth of the XCN band and the red component (rc) of CO-ice as proposed by Paper I.}
\label{rc4}
\end{figure} 
The CO-ice band, observed at around 2140\,cm$^{-1}$ toward all lines-of-sight here, has been analysed in detail in Paper\,I, using a phenomenological decomposition into 3 components to fit its band profile. The most red-centred component (rc) was attributed to CO in a hydrogen-bonded environment, although this cannot fully explain its extended red wing. It was suggested that the rc may evolve from pure CO-ice when thermal annealing of the ice induces CO to migrate into, and get trapped in, porous H$_2$O-ice. As such, this component may trace the thermal history of the interstellar ice mantle \citep[][Paper\,I]{Tielens1991}. Hence, the relationship between the optical depth of the rc of CO-ice and the XCN band may provide information on the thermal history of the XCN band as well. Paper\,I found a tentative correlation between this rc and the optical depth of the XCN band and suggested that this was actually due to a component of the XCN band, centred between 2170-2180\,cm$^{-1}$.\\
\indent Fig.~\ref{rc4}a presents the optical depth of the 2165.7-component versus the optical depth of the rc of the CO-ice band, showing that this component and the rc of CO-ice are not correlated. The shaded area marks the correlation suggested by Paper\,I. However, those sources that show a strong 2165.7-component also tend to have a strong CO rc, but not the other way around. \\
\indent In Fig.~\ref{rc4}b the optical depth of the 2175.4-component is plotted against the optical depth of the rc of the CO-ice band. In contrast to the 2165.7-component, the 2175.4-component matches the correlation indicated by the shaded area much better, particularly at rc optical depths of below 0.6, corroborating the idea posed by Paper I. However, the YSO's with deep rc optical depths (and deep 2165.7-component optical depths) fall outside the correlation proposed by Paper I. \\
\indent For completion, Fig.~\ref{rc4}c illustrates the relation between the integrated area of the XCN band and the optical depth of the rc of CO-ice, showing that most low-mass YSO's match the correlation suggested by Paper\,I (shaded area) and reflecting that in these objects the 2175.4-component dominates the XCN band (see Sect.~\ref{corh2o}). Conversely, approximately half of the XCN bands observed toward high-mass YSO's clearly exhibit larger integrated areas than would be expected from that correlation.
\section{Discussion}
\label{disc}
\subsection{The OCN$^-$ abundance}
\label{ocn_abun}
\begin{figure}
\centering
\includegraphics[width=8.5cm]{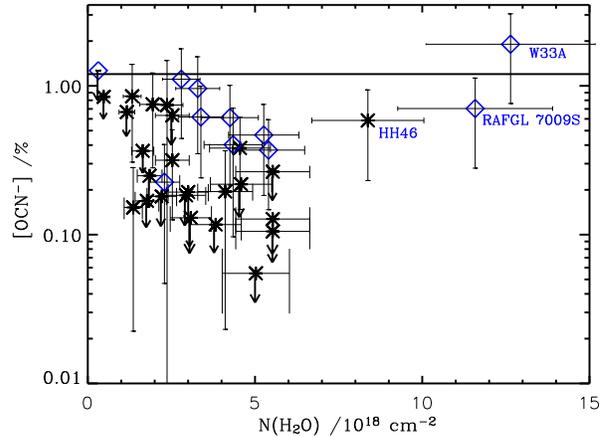}
\caption{The OCN$^-$ abundance, [OCN$^-$], plotted with respect to the H$_2$O column density, $N$(H$_2$O). High-mass objects are indicated by diamonds, all others, dominated by low-mass YSO's, by stars. 3\,$\sigma$ uncertainties are given,  or upper limits are indicated by arrows. The black horizontal line marks the 1.2\,\% maximum abundance of OCN$^-$ that may be formed photochemically under interstellar conditions, as determined from laboratory studies.}
\label{ocn4}
\end{figure} 
Based on the discussion in Sect.~\ref{ocnspec}, only the 2165.7-component of the astronomical spectra can be confidently assigned to OCN$^-$. Thus, the OCN$^-$ column density, $N$(OCN$^-$), and its abundance with respect to H$_2$O, [OCN$^-$], are derived from the optical depth of the 2165.7-component via:
\begin{equation}
{\rm [OCN^-]}={\frac{N({\rm OCN^-})}{N({\rm H_2O})}}={\frac{{\int{\tau_{\rm 2165.7}\,d\nu}\times{A_{\rm OCN^-}^{-1}}}}{N({\rm H_2O})}}
\end{equation}
where the FWHM of $\tau_{\rm 2165.7}$ is fixed at 26\,cm$^{-1}$ and
$A_{\rm OCN^-}$ is 1.3$\times$10$^{-16}$ cm\,molec.$^{-1}$ \citep[][see
Sect.~\ref{bandstrength} for a discussion on the value of $A_{\rm
OCN^-}$]{vanBroekhuizen2004}. The inferred $N$(OCN$^-$) and [OCN$^-$]
are summarised in Table~\ref{fitresults}. 

In Fig.~\ref{ocn4} the OCN$^-$ abundance with respect to $N$(H$_2$O),
[OCN$^-$], is plotted against $N$(H$_2$O). All sources studied here
show abundances of $\leq$1.0\,\%, except for W33\,A. Excluding the
high-mass stars, gives abundances of $\leq$\,0.85\,\%. The high
signal-to-noise of the spectra allows the detection of the XCN band
down to optical depths of 0.01. As a result, very strict upper limits
have been put on the OCN$^-$ abundance along some lines of sight. 

\begin{table*}
\caption{Results of the XCN band fitting procedure}
\centering
\label{fitresults}
\tiny
\begin{tabular}{l|ll|l|l|l}
\hline
\hline
Source\,$^a$\,\,\,\,\,\,\,\,\,\,\,\,\,\,\,\,\, \,\,\,\,\,\,\,\,\,\,\,\,\,\,\,\,\,   &      \multicolumn{1}{c}{$\tau$(2165.7)$^{\it b}$}        &         \multicolumn{1}{c}{$\tau$(2175.4)$^{\it b}$}      &         \multicolumn{1}{|c}{$N$(OCN$^-$)$^{\it c}$}& \multicolumn{1}{|c}{[OCN$^-$]$^{\it c}$} & \multicolumn{1}{|c}{$\tau$(H$_2$O)\,$^{\it d}$}\\
&2165.7 cm$^{-1}$ &2175.4 cm$^{-1}$ & $\times$10$^{16}$\,molec.\,cm$^{-2}$& \multicolumn{1}{c|}{\%} &  \\
\hline
\multicolumn{6}{l}{\bf $\rho$ Ophiuchus}\\
\hline
IRS42            &$\leq$0.023     &0.020$\pm$0.010 &$\leq$0.46   &$\leq$0.25 & 1.17\\
IRS43            &$\leq$0.02     &0.072$\pm$0.030 &$\leq$0.40    &$\leq$0.18 & 1.40\\
IRS44            &$\leq$0.02     &$\leq$0.024     &$\leq$0.40   &$\leq$0.13 & 1.95\\
IRS46            &$\leq$0.039     &$\leq$0.035     &$\leq$0.78   &$\leq$0.68 & 0.74\\
IRS51            &0.073$\pm$0.013 &0.037$\pm$0.015 &1.46$\pm$0.26&0.76$\pm$0.12& 1.23\\
IRS63            &$\leq$0.015     &0.075$\pm$0.016 &$\leq$0.3    &$\leq$0.17 & 1.12\\
WL12             &-               &0.062$\pm$0.032 &-            &-      & 1.95\\
CRBR 2422.8      &$\leq$0.029     &0.120$\pm$0.045 &$\leq$0.58   &$\leq$0.19 & 1.90\\
Elias 32         &$\leq$0.030     &0.074$\pm$0.036 &$\leq$0.60   &$\leq$0.36 & 1.04\\
VSSG17           & -              &$\leq$0.095     & -           &- & 0.87\\
RNO 91           &0.087$\pm$0.010 &0.066$\pm$0.017 &1.74$\pm$0.2 &0.40$\pm$0.09 & 2.88\\
\hline
\multicolumn{6}{l}{\bf Serpens}\\
\hline
EC90 A           &0.010$\pm$0.006 &0.039$\pm$0.008 &0.2$\pm$0.12 &0.15$\pm$0.05 & 0.86\\
EC90 B           &0.056$\pm$0.012 &0.030$\pm$0.007    &1.12$\pm$0.24&0.86$\pm$0.12 & 0.84\\
EC82             &$\leq$0.02     &$\leq$0.026     &$\leq$0.40   &$\leq$0.86 & 0.30\\
CK2              &$\leq$0.027     &$\leq$0.089     &$\leq$0.54  &$\leq$0.19 & 1.86$\pm$0.19\,$^{\it e}$\\
SVS 4-9          &$\leq$0.08     &$\leq$0.04  &$\leq$1.60   &$\leq$0.64 & 1.6$\pm$0.1\,$^{\it f}$\\
SVS 4-5          &$\leq$0.073     &$\leq$0.010     &$\leq$1.46  &$\leq$0.27 & 3.50$\pm$0.30\,$^{\it f}$\\
\hline
\multicolumn{6}{l}{\bf Orion}\\
\hline
TPSC 78          &0.155$\pm$0.003  &$\leq$0.025    &3.1$\pm$0.06 &1.12$\pm$0.14 & 1.77\\
TPSC 1           &0.158$\pm$0.032  &$\leq$0.042    &3.16$\pm$0.64&0.97$\pm$0.13 & 2.08\\
Reipurth 50      &0.100$\pm$0.005    &0.049$\pm$0.012&2.0$\pm$0.1  &0.38$\pm$0.08 & 3.42\\
\hline
\multicolumn{6}{l}{\bf Corona Australis}\\
\hline
HH100 IRS        &0.04$\pm$0.002   &0.098$\pm$0.003 &0.8$\pm$0.04 &0.32$\pm$0.08 & 1.6 \\
HH46             &0.245$\pm$0.019  &0.095$\pm$0.032 &4.9$\pm$0.38 &0.59$\pm$0.01 & 5.3$\pm$3.0\,$^{\it g}$\\
RCrA IRS7A       &$\leq$0.029      &$\leq$0.074     &$\leq$0.58   &$\leq$0.11 & 3.5\\
RCrA IRS7B       &$\leq$0.035      &$\leq$0.050     &$\leq$0.70    &$\leq$0.13 & 3.5\\
RCrA IRS5A       &0.040$\pm$0.025  &0.123$\pm$0.036 &0.80$\pm$0.5  &0.20$\pm$0.06 & 2.60\\
RCrA IRS5B       &$\leq$0.022      &$\leq$0.054     &$\leq$0.44   &$\leq$0.12 & 2.42\\
\hline
\multicolumn{6}{l}{\bf Chamaeleon}\\
\hline
ChaIRN           &0.088$\pm$0.069   &$\leq$0.055    &1.76$\pm$1.38&0.75$\pm$0.12 & 1.5\\
ChaIRS 6A        &$\leq$0.05        &$\leq$0.108    &$\leq$1.0    &$\leq$0.22 & 2.90\\
\hline
\multicolumn{6}{l}{\bf Vela}\\
\hline
LLN17           &0.104$\pm$0.011   &0.048$\pm$0.014 &2.08$\pm$0.22&0.62$\pm$0.11 & 2.14\\
LLN20           &0.087$\pm$0.040   &$\leq$0.024     &1.74$\pm$0.8 &0.41$\pm$0.09 & 2.75\\
LLN33           &0.130$\pm$0.032   &$\leq$0.051     &2.6$\pm$0.64 &0.62$\pm$0.11 & 2.69\\
LLN39           &$\leq$0.02        &$\leq$0.02      &$\leq$0.40   &$\leq$1.3     & 0.20\\
LLN47           &-                 &0.050$\pm$0.033 &-            & -& 1.15\\
\hline
\multicolumn{6}{l}{\bf Taurus}\\
\hline
LDN 1489 IRS    &$\leq$0.014       &0.070$\pm$0.016 &$\leq$0.28   &$\leq$0.06 & 3.18\\
\hline
\multicolumn{6}{l}{\bf{Additional sources}\,$^{\it h}$}\\
\hline
RAFGL 2136       &0.123$\pm$0.011   &$\leq$0.051     &2.46$\pm$0.22&0.47$\pm$0.09 & 3.33\,$^{\it i}$\\
NGC7538 IRS1    &-                 &$\leq$0.019     &-            &- & 1.47\\
RAFGL 7009S     &0.408$\pm$0.021   &0.104$\pm$0.026 &8.16$\pm$0.42&0.71$\pm$0.11 & 7.33\,$^{\it j}$\\
W33A            &1.204$\pm$0.030    &$\leq$0.004     &24.08$\pm$0.60&1.93$\pm$0.01 & 8.00$\pm$0.75\,$^{\it k}$\\
RAFGL 989        &0.026$\pm$0.013   &0.019$\pm$0.017 &0.52$\pm$0.26&0.23$\pm$0.06 & 1.45\\
\hline
\end{tabular}
\begin{itemize}
\item[]\footnotesize{$^{a}$ Source spectra taken from Paper I, unless otherwise stated. $^{b}$ 3$\sigma$ uncertainties or upper limits are given. $^{c}$ Column densities and abundances derived from the 2165.7-component using Eq.2. $^{d}$ H$_2$O optical depths from Dartois et al. private communication, unless otherwise stated, and derived according to Sect.~\ref{observations}. Uncertainties are between 10 to 20\,\%. $^{e}$ \citet{Eiroa1989} $^{f}$ \citet{Pontoppidan2004} $^{g}$ \citet{Boogert2004} $^{h}$ ISO-data archive. $^{i}$ \citet{Gerakines1999} $^{j}$ \citet{Dartois1999} $^{k}$ \citet{Gibb2000}}
\end{itemize}
\end{table*}

\subsection{OCN$^-$ abundance variations}
The sample of XCN observations presented here is large enough to obtain
statistically significant information on the distribution of
abundances and to compare between different classes of cloud
environments, such as low- and high-mass star-forming regions. As seen
in Fig.~\ref{ocn4} and Table~\ref{fitresults}, the abundance of
OCN$^-$ is observed to vary by at least a factor of 10-20 among the
lines of sight in the sample. This observed range of OCN$^{-}$
abundances is larger than that of CH$_3$OH and second only to that of ice
species whose abundances are sensitive to low-temperature freeze-out
and evaporation, such as CO.\\
\indent A particularly interesting point is that
some of the observed lines of sight toward
low-mass sources are located in close proximity to each other in the
plane of the sky (100-10\,000\,AU). Such objects provide a direct
probe of the physical size of regions with enhanced (or reduced) OCN$^-$
abundances. In particular, several close binaries in the sample probe
variations in abundances on scales of a few hundred AU, i.e. at the
scale of circumstellar disks. 
Some of the best upper limits on the OCN$^-$ abundance of $<0.1\,\%$
are observed toward IRS 43 and IRS 44 in the $\rho$ Ophiuchus
star-forming cloud. These lines of sight are located approximately
$4\arcmin$ or 30\,000\,AU from IRS 51, which has a measured OCN$^-$
abundance of 0.8\,\%. This shows that OCN$^-$ is formed in localised
regions even within the same low-mass star-forming cloud complex. The
individual components of the close binary source EC 90A+B, which were
observed separately, appear to have OCN$^-$ abundances varying by a
factor of 6. Since the separation between the two components is only
1\farcs6, corresponding to approximately 400\,AU, this is an
indication that processes occurring on the size scale of circumstellar
disks may be required to form OCN$^-$ in large quantities. Note,
however, the spectrum observed toward EC 90B is complicated by complex
emission and absorption from gaseous CO, so that the observed
difference in OCN$^-$ abundances for this binary requires further
investigation. The two components in another close binary, RCrA IRS5,
also show a tentative difference in OCN$^-$ abundance of a factor of
2. Clearly, only slightly more sensitive observations of embedded
close binaries will be able to confirm the observation of a varying
OCN$^-$ abundance on scales of a few hundred AU.
\subsection{OCN$^-$ formation toward low-mass YSOs}
The observed abundances and their variations for our large sample of
sources put important constraints on the formation mechanism of
OCN$^-$.  All inferred OCN$^-$ abundances presented here allow
quantitatively for a photochemical formation mechanism (see
Sect.~\ref{ocnform}).  Only W33\,A, an embedded O-type star with
a luminosity of $10^{5}\,L_{\odot}$ \citep{Mueller2002} with an
OCN$^-$ abundance of $\sim$\,2\% is at the edge of the range, as are
the Galactic center source GC:IRS19 and the external galaxy NGC 4945
presented elsewhere (see van Broekhuizen et al.\ 2004).  If we define
high- and intermediate mass YSO's as those more luminous than
50\,$L_{\odot}$, it is found that OCN$^-$ abundances of order
0.85--1\,\% are observed toward high- intermediate- and low-mass stars
in our sample, but strict upper limit abundances of $<0.2\,\%$ are
observed only toward low-mass stars. This does indicate that some
differences exist in the efficiency of OCN$^-$ production between
high- and low-mass star-forming regions, and may suggest that UV
photons play a role.  However, comparison with laboratory studies
shows that the photon dose required to produce the highest OCN$^-$
abundances (of $\sim$\,0.85\,\%) detected here toward low-mass YSO's
are $\sim$\,40\,$\times$ larger than can be accounted for by the
cosmic ray induced photons that any ice might receive within the
typical lifetime of the pre- and protostellar phases (see
Sect.~\ref{ocnform}). Adopting the somewhat higher cosmic-ray induced
photon fluxes estimated by \citet{Shen2004} would lower these
timescales by an order of magnitude, and they could be further decreased
by a factor of 2 if the initial NH$_3$ abundance were as high as
30\,\%, but it remains difficult to fully reconcile the ages.\\
\indent Another, perhaps more serious problem with the UV photoproduction
scenario is that little variation in the OCN$^-$ abundance from source
to source is expected if the cosmic-ray induced photon flux is assumed
to be the same everywhere. Thus, local sources of radiation ---either
UV from the young star or star-disk boundary--- need to be invoked to
explain the observed small-scale variations. A well-known concern is
whether any of this UV radiation is able to reach the bulk of the ices
because of the large extinctions involved.  Spherically-symmetric
models of YSO envelopes have been constructed, taking the variation of
the density, dust temperature and UV radiation with depth into account
\citep[e.g.][]{Jorgensen2002}, and it has been shown that any UV from
the star only reaches the inner few \% of the envelope mass
\citep[e.g.,][]{Stauber2004}. Thus, special geometries are required
to allow the UV to escape and reach the bulk of the ice, such as
outflow cavities or flaring turbulent disks in which the ices are
cycled from the shielded midplane to the upper layers.  One
alternative option is UV radiation created by secondary electrons
resulting from X-rays from the young star. Model calculations show
that the X-rays, and thus the UV radiation, penetrate much deeper into
the envelope than the UV from the star itself so that it can reach a
larger fraction of the ices \citep[][]{Stauber2005}. It would be
worthwhile to couple such models with an ice chemistry to check the
OCN$^-$ production quantitatively, but this is beyond the scope of
this paper.\\
\indent An alternative scenario for OCN$^-$ formation is through thermal
heating of HNCO in the presence of NH$_3$- or H$_2$O-ices. Assuming a
15--100\,\% efficiency of this mechanism as determined under
laboratory conditions (see Sect.~\ref{ocnform}), the OCN$^-$
abundances toward low-mass YSO's presented here imply that 0.85--5\,\%
HNCO should be present in the solid state prior to the formation of
OCN$^-$ to explain the highest observed abundances of
$\sim$0.85\,\%. To date, HNCO has been detected in interstellar gas
associated with star-forming regions, with column densities in the
range from 0.3--7.5$\times$10$^{14}$\,cm$^{-2}$
\citep{Wang2004,Zinchenko2000,Kuan1996,Jackson1984,vanDishoeck1995}. This
is a factor of 100 less than what is expected from the OCN$^-$ column
densities derived here, assuming all the OCN$^-$ evaporates as
HNCO. In the solid state, only HNCO upper limits, of the order of 0.5
to 0.7\,\% with respect to H$_2$O-ice, are determined toward W\,33A,
NGC\,7538 IRS9 and AFGL2136 \citep{vanBroekhuizen2004}. Toward the
low-mass YSO Elias\,29, this is $\leq$\,0.3\,\% \citep[derived
from][]{Boogert2002}. The ease with which HNCO is deprotonated in a
H$_2$O-rich environment implies that HNCO would be difficult to detect
toward regions where the ice is warmer than 15\,K, although its
presence may be observable by means of its 4.42\,$\mu$m feature toward
the coldest regions where the ice temperature is close to 10\,K. On
the basis of $\tau_{\rm OCN^-}$ derived here, and the band strength of
HNCO \citep{vanBroekhuizen2004}, $\tau_{\rm HNCO}$ is expected to
range from 0.001 to 0.08, a factor of 5 less than observed for
OCN$^-$.\\
\indent Chemical models predict that HNCO is formed on interstellar grains via
grain-surface reactions with an abundance upper limit of
$\leq$\,1--3\,\% with respect to H$_2$O-ice
\citep{Hasegawa1993,Keane2001}. Given the observed upper limits and
model abundances for HNCO, thermal processing also has difficulties to
explain the highest observed OCN$^-$ abundances, but it is a serious
alternative to UV-photoprocessing for the formation of at least a
fraction of OCN$^-$ in interstellar ices.  The observed OCN$^-$
abundance variations would in this scenario have to result from
variations in the grain-surface chemistry rates forming the reactants
that lead to OCN$^-$, for example due to temperature or availability
of atomic N.  \\
\indent On the basis of its spectroscopy, it is not possible to draw a
distinction between OCN$^-$ formation via proton-, electron-, UV- or
thermally induced processes, neither from laboratory spectra
(Sect.~\ref{ocnspec}), nor from astronomical observations of the $\rm
2165.7\,cm^{-1}$ OCN$^-$ band. 
\section{Conclusion}
\indent The XCN band has been observed toward 39 YSO's enabling the
first detailed study of the OCN$^-$ abundance in a large sample of
low-mass YSO's. Statistical analysis of the band centre position
distribution proves unequivocally that the band contains at
least two components, of which only one (centred at 2165.7\,cm$^{-1}$)
can be associated with OCN$^-$ as its carrier, based on laboratory
OCN$^-$ spectra. We conclude that in all cases the OCN$^-$ is embedded
in a strongly hydrogen-bonded, and possibly thermally annealed, ice
environment. A phenomenological decomposition was undertaken to fit
the full XCN band profile toward each line of sight using two
components, centred at 2165.7\,cm$^{-1}$ (FWHM = 26\,cm$^{-1}$) and
2175.4\,cm$^{-1}$ (FWHM = 15\,cm$^{-1}$). OCN$^-$ abundances were
derived from the 2165.7-component of this fit. 

Typically, OCN$^-$ is detected toward the low-mass YSO's in our sample
at abundances $\leq$\,0.85\,\% (with respect to the H$_2$O-ice column
density) and $\leq$\,1.0\,\% toward high-mass YSO's, with the
exception of W33\,A. A large dynamic range of abundances is observed,
varying by at least a factor of 10--20. Together with the variations
in abundance observed between regions separated by only 400 AU, this
provides a strong indication that OCN$^-$ formation is a highly
localised process.  The inferred OCN$^-$ abundances
allow quantitatively for a photochemical formation mechanism, but the
observed large variations are more difficult to explain unless local
sources of UV radiation or X-rays are invoked which reach a large
fraction of the ices. Alternatively, OCN$^-$ can be formed in the bulk
of the ice from the solvation of HNCO, which may have formed via
surface reactions. However, this too has difficulties. Further diagnostics are required to be more
conclusive about the chemistry leading to OCN$^-$ formation in
interstellar environments. These include the construction of more
detailed physical models of the sources presented here to derive the
effects of varying grain temperatures and UV radiation fields on the
different formation mechanisms of OCN$^-$, slightly more sensitive
observations to probe OCN$^-$ abundance variations on a 100 AU scale,
and laboratory data on the formation of HNCO via surface reactions
at various temperatures.
\begin{acknowledgements}
This research was financially supported by the Netherlands Research
School for Astronomy (NOVA) and a NWO Spinoza grant. Thanks to
F. Lahuis, E. Dartois, L. d'Hendecourt, A.G.G.M. Tielens, W.-F. Thi
and S. Schlemmer for many useful discussions on the VLT program, and
to an unknown referee for useful comments that led to considerable
improvements to the article.
\end{acknowledgements}
\bibliography{aamnem99,1711}
\bibliographystyle{aa}
\end{document}